\def\sqr#1#2{{\vcenter{\vbox{\hrule height.#2pt\hbox{\vrule
width.#2pt height#1pt \kern#1pt\vrule width.#2pt}\hrule height.#2pt}}}}
\def\square{\mathchoice\sqr54\sqr54\sqr{2.1}3\sqr{1.5}3}

\magnification\magstep1
\font\cst=cmr10 scaled \magstep3
\font\csc=cmr10 scaled \magstep2
\vglue 1.5cm
\centerline{\cst  Gravity on branes}
\vskip 1.5 true cm
\centerline{Nathalie Deruelle$^{1,2,3}$  and Joseph Katz$^{4}$}
\vskip 1 true cm
\centerline{$^1$\it D\'epartement d'Astrophysique Relativiste et de
Cosmologie,}
\centerline{\it UMR 8629 du Centre National de la Recherche Scientifique,}
\centerline{\it Observatoire de Paris, 92195 Meudon, France}
\medskip
\centerline{$^2$ \it Institut des Hautes Etudes Scientifiques,}
\centerline{\it 91440, Bures-sur-Yvette, France}
\medskip
\centerline{$^3$ \it Centre for  Mathematical Sciences, DAMTP, University of Cambridge,}
\centerline{\it Wilberforce Road, Cambridge, CB3 0WA, England}
\medskip
\centerline{$^4$ \it Racah Institute of Physics, Hebrew University}
\centerline{\it 91904, Jerusalem, Israel.}

\vskip 0.5cm
\centerline{April 2nd 2001}

\vskip 0.5cm
{\bf Pacs Numbers}~: 98.80.Cq, 98.70.Vc

\vskip 1cm
\noindent
{\bf Abstract}
\bigskip
We consider the four dimensional discontinuity generated by two
identical pieces of a five dimensional space pasted along their edge 
(that is a ``brane" in a
``$Z_2$-symmetric"  ``bulk"). Using a four plus one decomposition of the Riemann tensor we write the equations
for gravity on the brane and recover in a simple manner a number of known ``brane world" scenarios. We study under
which conditions these equations reduce, exactly or approximately, to the four dimensional Einstein equations. We
conclude that if the bulk is imposed to be only an Einstein space near the brane, Einstein's equations can be recovered
approximately on the brane, but if it is imposed to be strictly anti-de Sitter space then the Einstein equations cannot
hold, even approximately, on a quasi-Minkowskian brane, unless matter obeys a very contrived equation of state.

\vfill\eject

\noindent
{\csc I. Introduction}
\medskip
Since their covariant description by Israel (clarifying earlier treatments by e.g. Lanczos and Darmois)
 [1], thin shells have been intensively used as a model for matter in general relativity. However a
question is rarely asked when considering a four dimensional spacetime, to wit whether or not matter on the shell obeys
the three dimensional Einstein equations, as there is no (experimental) reason why it should.

The situation changed recently with the increasing interest in gravity theories within spacetimes with large
extra dimensions and the advent of the idea that our universe may be a four dimensional singular hypersurface, or
``brane", in a five dimensional spacetime, or ``bulk" [2]. Indeed, in this new context, it becomes crucial to recover
Einstein gravity for realistic matter on the brane, at least to some approximation compatible with the present
experiments. 

The Randall Sundrum scenario [3], where our  universe is a four dimensional quasi-minkowskian edge of a
double-sided perturbed anti-de Sitter spacetime, was the first explicit model where the linearized Einstein equations
were claimed to hold on the brane. This claim was substantiated by further analyses and the corrections to Newton's law
calculated  [4]. Soon followed the building of cosmological models where the brane is taken to be a Robertson-Walker
spacetime which can tend at late times to the standard big bang scenario [5-6]. The perturbations of the geometry and
matter content of these models, in the
view of calculating the microwave background anisotropies, are currently being studied  [7]. More sophisticated
models, including, for example,  two branes or curvature squared corrections in the bulk gravity equations are also being
considered [8-9].

In most papers the issue of whether or not Einstein's equations can be recovered on the brane is slightly
confused for the following reason~: some authors take a ``brane world" point of view, that is they ignore as much as they
can the bulk, which shows up in the gravity equations on the brane as some extra radiation fluid or seeds~; whereas other
authors impose a geometry for the bulk (to be e.g. perturbed anti-de Sitter spacetime) and see how this geometry
influences the equations for gravity in the brane. This divide can be seen on the techniques used~: the first category of
authors tends to use gaussian normal coordinates which are well adapted to the brane, whereas the second
tends to use coordinates adapted to the bulk (e.g. conformally minkowskian or Schwarzschild-like coordinates). In this
paper we shall adopt the first point of view. We make the junction with the ``bulk" point of view in an accompanying
paper [10].

The issue however can be described in a coordinate independent way as follows.  Start with a (N+1)-dimen\-sional
``generating" spacetime ${\cal  V}_{(N+1)}$ (that one can visualize as a surface embedded in a higher dimensional
space).  Assume that ${\cal V}_{(N+1)}$ satisfies Einstein's equations, that is that
the Einstein tensor  of ${\cal V}_{(N+1)}$ is linearly related to some (smooth) stress-energy tensor ${\cal T}_{AB}$.
Consider in
 ${\cal V}_{(N+1)}$ a N-dimensional hypersurface ${\cal M}_N$. Cut ${\cal V}_{(N+1)}$ along
${\cal M}_N$ into two parts ${\cal V}^1_{(N+1)}$ and ${\cal V}^2_{(N+1)}$  and keep, say ${\cal V}^1_{(N+1)}$, which
has now a boundary ${\cal M}_N$. Then make a copy of ${\cal V}^1_{(N+1)}$ and paste these two identical pieces
 along  ${\cal
M}_N$~; call the new spacetime with a discontinuity ${\cal M}_{(N+1)}$. 

This cutting,
copying and pasting procedure  is the geometrical expression of the so-called $Z_2$-symmetry
\footnote *{There are two distinct ways of pasting
two identical pieces together along a cut: one
into another (double sided space) and one to another (single sided space).
There is no way to distinguish these
two constructs by the intrinsic and extrinsic curvatures  of their
discontinuity because they are the same for double sides
or single sided spaces.}. 
In brane cosmology
language, ${\cal M}_{(N+1)}$ is the bulk and ${\cal M}_N$ is the brane. Since ${\cal M}_{(N+1)}$ has a delta-like curvature
singularity at its edge ${\cal M}_N$,  it can satisfy Einstein's equations only if a delta-like tensor is added to ${\cal
T}_{AB}$. This new tensor (which can be visualized as the ``glue" necessary to paste the two copies of
${\cal V}^1_{(N+1)}$) is the stress-energy tensor of matter in the brane ${\cal M}_N$. It is given  by the integrated
Einstein equations across  ${\cal M}_N$, also called the Lanczos-Darmois-Israel equation, and is linearly related to the
extrinsic curvature of ${\cal M}_N$ in the generating space ${\cal V}_{(N+1)}$.

A first remark is that any brane in any bulk cannot be kept as a candidate to represent our universe since
conditions (such as energy conditions or an equation of state) must be imposed on the stress-energy tensor of the brane,
that is, through the Lanczos-Darmois-Israel  equation, on its extrinsic curvature. 

The conditions on the bulk and the
brane are even more stringent if we impose that the brane ${\cal M}_N$ itself satisfies Einstein's equations (or an
approximate version of those) because this implies highly non trivial
relations between the
extrinsic  and  intrinsic curvatures of the brane.

In this paper we show how the equations for gravity on a brane depend only on the geometry of the bulk {\it near} the
brane, but do so crucially  (we shall be more precise about what we mean by ``near" below). We will first see that if the
geometry of the bulk can be chosen at will near the brane, then the Einstein equations can always be recovered on the
brane, whatever the matter we choose on it. Second, we will see that if the bulk is imposed to be an Einstein
space near the brane then the Einstein equations can also be recovered on the brane, under the condition however that
terms quadratic in the stress-energy tensor of matter can be neglected (this result is already well known in the context
of brane cosmologies [5-6])~; otherwise matter must satisfy a very special equation of state (typically
$P=-{1\over3}\rho$). Finally we will see that if the bulk is imposed to be maximally symmetric near the brane, then the
Einstein equations cannot in general be recovered on the brane, even when terms quadratic in the stress-energy tensor of
matter can be neglected (an exception being the case when the brane is a Robertson-Walker spacetime). In particular, we
will see that the linearized Einstein equations cannot hold on a quasi-Minkowskian brane at the edge of a strictly anti-de
Sitter bulk, unless matter obeys a very contrived equation of state. Our approach, which is based on a four plus one
decomposition of the bulk Riemann tensor and an identification of the extrinsic curvature of the brane with its
stress-energy tensor (thanks to the Lanczos-Darmois-Israel equations), is similar to that of references [11]
and our results are an extension of those presented there.

The paper is organised as follows~: in section 2 we express, in gaussian normal coordinates, the metric near the brane
in terms of the stress-energy tensor of matter on the brane and an extra, ``seed" tensor. We also write the
equations for gravity on the brane in terms of these two tensors and show that if the geometry of the bulk
near the brane can be chosen at will then the exact Einstein equations can be recovered on the brane. In section 3 we
restrict our attention to bulks which are Einstein spaces near the brane and in section 4 to maximally symmetric bulks.
Section 5 draws a few conclusions and we relagate to an appendix the procedure to obtain by iteration the metric in the
whole bulk, when matter in the bulk is known everywhere and not only near the brane.

\bigskip
\noindent
{\csc 2. The equations for gravity on a brane}
\bigskip
Consider an arbitrary smooth five dimensional ``generating" space ${\cal V}_5$ that we foliate (at least locally) by a
family of timelike hypersurfaces
$\Sigma_y$. In gaussian normal coordinates the metric of ${\cal V}_5$ reads 
$$ds^2\equiv \gamma_{AB}dx^Adx^B=dy^2+\gamma_{\mu\nu}(x^\rho,y)dx^\mu dx^\nu\eqno(2.1)$$
where $x^\rho$ are four coordinates (one timelike, three spacelike) parametrizing the  hypersurfaces $\Sigma_y$
 and where $x^5\equiv y=Const.$ are the equations of  $\Sigma_y$. We introduce the extrinsic curvature of an
hypersurface $\Sigma_y$  and its trace
$${\cal K}_{\mu\nu}\equiv -{1\over2}{\partial\gamma_{\mu\nu}\over\partial y}\qquad,\qquad {\cal
K}\equiv\gamma^{\rho\sigma}{\cal K}_{\rho\sigma}\eqno(2.2)$$ as well as the Lanczos tensor
$${\cal L}_{\mu\nu}\equiv {\cal K}_{\mu\nu}-\gamma_{\mu\nu}{\cal K}\eqno(2.3)$$
that we decompose in terms of  a ``$\tau$-tensor" as
$${\cal L}_{\mu\nu}\equiv{1\over2}\lambda\gamma_{\mu\nu}+{\kappa\over2}\tau_{\mu\nu}\,,\eqno(2.4)$$
$\lambda$ being a ``tension" and  $\kappa$ a coupling constant.

 We now single out the hypersurface
$\Sigma_0\equiv{\cal M}_4$.  Near ${\cal M}_4$ the metric can be expanded in Taylor series as
$$\gamma_{\mu\nu}(x^\rho,y)=g_{\mu\nu}(x^\rho)+k_{\mu\nu}(x^\rho)\,y+{1\over2}l_{\mu\nu}(x^\rho)\,y^2+{\cal
O}(y^3)\eqno(2.5)$$
where $g_{\mu\nu}$ is the metric on  ${\cal M}_4$.
We write he expansion of the $\tau$-tensor as
$$\tau_{\mu\nu}=T_{\mu\nu}(x^\rho)+\Theta_{\mu\nu}(x^\rho)\,y+{\cal O}(y^2)\eqno(2.6)$$
where $T_{\mu\nu}$ and $\Theta_{\mu\nu}$ can be expressed in terms of $k_{\mu\nu}$ and $l_{\mu\nu}$ (see
Appendix). Conversely
$k_{\mu\nu}$ and $l_{\mu\nu}$ can be expressed in terms of $T_{\mu\nu}$ and $\Theta_{\mu\nu}$ so that the metric
near the brane can be written as (1) with
$$\eqalign{\gamma_{\mu\nu}=& g_{\mu\nu}\left(1+{1\over3}\lambda\,y+{1\over18}\lambda^2\,y^2\right)-
\kappa y\left(1+{1\over6}\lambda\,y\right)\left(T_{\mu\nu}-{1\over3}T\,g_{\mu\nu}\right)\cr&
-{1\over2}\kappa y^2\left[\left(\Theta_{\mu\nu}-{1\over3}\Theta
g_{\mu\nu}\right)-{1\over3}\kappa\left(g_{\mu\nu}T_{\rho\sigma}T^{\rho\sigma}-TT_{\mu\nu}\right)\right]+ {\cal
O}(y^3)\,\cr}\eqno(2.7)$$
traces being defined by means of $g^{\mu\nu}$.
\vfill\eject
\bigskip

We now assume that ${\cal V}_5$ satisfies the five dimensional Einstein equations
$${\cal G}_{AB}={1\over6}\lambda^2\gamma_{AB}+\kappa {\cal T}_{AB}\eqno(2.8)$$
where ${\cal G}_{AB}$ is its Einstein tensor,  and ${\cal
T}_{AB}$ a smooth stress-energy tensor.  If we use the standard four plus
one decomposition of the five dimensional Riemann tensor ${\cal R}_{ABCD}$
$$\eqalign{{\cal R}_{y\mu y\nu}&={\partial\over\partial y}{\cal K}_{\mu\nu}+{\cal K}_{\rho\mu}{\cal K}^\rho_\nu\cr
{\cal R}_{y\mu \nu\rho}&=\nabla_\nu{\cal K}_{\mu\rho}-\nabla_\rho{\cal K}_{\mu\nu}\cr
{\cal R}_{\mu \nu\rho\sigma}&=^4{\cal R}_{\mu \nu\rho\sigma} +{\cal K}_{\mu\sigma}{\cal K}_{\nu\rho}
-{\cal K}_{\mu\rho}{\cal K}_{\nu\sigma}\cr}\eqno(2.9)$$
where $\nabla_\mu$ and $^4{\cal R}_{\mu \nu\rho\sigma}$ are the covariant derivative and Riemann tensor associated
with the metric $\gamma_{\mu\nu}(x^\rho,y)|_{y=Const.}$, we can rewrite the five dimensional Einstein equations (8) in
terms of the  quantities introduced previously, at zeroth order in $y$, that is on ${\cal M}_4$, as
$$\eqalign{G_{\mu\nu}&=-{\kappa\lambda\over6}T_{\mu\nu}-{\kappa\over2}\Theta_{\mu\nu}-\cr
&\qquad -{\kappa^2\over2}
\left[T_{\mu\rho}T^\rho_\nu-{1\over6}TT_{\mu\nu}+{g_{\mu\nu}\over4}
\left(T_{\rho\sigma}T^{\rho\sigma}-{1\over3}T^2\right)\right]+\kappa {\cal
T}_{\mu\nu}|_{y=0}\cr}\eqno(2.10a)$$
$$D_\nu T^\nu_\mu=-2{\cal T}_{5\mu}|_{y=0}\eqno(2.10b)$$
$$-R=-{\kappa\lambda\over6}T +{\kappa^2\over4}\left(T_{\rho\sigma}T^{\rho\sigma}-{1\over3}T^2\right) +2\kappa
{\cal T}_{55}|_{y=0}\eqno(2.10c)$$
where $G_{\mu\nu}$ is the Einstein tensor of the metric $g_{\mu\nu}$, $(-R)$ its trace, and $D_\mu$ the covariant
derivative associated with $g_{\mu\nu}$. A consequence of (10) is 
$$\Theta-2{\cal T}^\rho_\rho+4{\cal T}_{55}=\kappa\left(-{5\over2}T_{\rho\sigma}T^{\rho\sigma}+
{2\over3}T^2\right)\,.\eqno(2.11)$$
Equations (10) can be seen as an ``initial value" problem (with inverted commas because ${\cal M}_4$ is a timelike
hypersurface)~: given a metric and its first derivative in $y$ on ${\cal M}_4$, that is,  with the notations used here, given
a metric
$g_{\mu\nu}$ on
${\cal M}_4$ and a  tensor
$T_{\mu\nu}$, satisfying the constraints (10b) and (10c), then equation (10a) gives  $\Theta_{\mu\nu}$ that is the
second $y$-derivative of the metric on the hypersurface. One then knows the metric and its first derivative on a
neighbouring hypersurface $y=\epsilon$ and, by iteration, one can get in principle the metric in the whole spacetime if
${\cal T}_{AB}$ is known everywhere (see the Appendix  for an illustration of such a procedure).

The reason for decomposing the five dimensional Einstein equations (8) in terms of $T_{\mu\nu}$ rather than the
extrinsic curvature of
${\cal M}_4$ as is usual, is that, in brane cosmology,  $T_{\mu\nu}$ is the stress-energy tensor of ordinary 
matter on the brane. Indeed,
as recalled in the
Introduction, the ``bulk" ${\cal M}_5$ is obtained by cutting ${\cal V}_5$ into two pieces along ${\cal M}_4$, by making a
copy of the $y\geq0$ piece, say, and pasting it along ${\cal M}_4$, which hence becomes a singular hypersurface, or
``brane".  The metric of ${\cal M}_5$ is continuous across ${\cal M}_4$ and reads
$$\eqalign{\bar\gamma_{AB}&=\gamma_{AB}(x^\rho,y)\qquad\hbox{for}\qquad y\geq0\cr
\bar\gamma_{AB}&=\gamma_{AB}(x^\rho,-y)\,\,\quad\hbox{for}\qquad y \leq0\,.\cr}\eqno(2.12)$$
The stress-energy tensor $\bar{\cal T}_{AB}$ is defined similarly.
The extrinsic curvature of ${\cal M}_4$ in ${\cal M}_5$ is $-(k_{\mu\nu}/2)$ when $y\to0_+$ and $+(k_{\mu\nu}/2)$
when
$y\to0_-$. The Einstein tensor of ${\cal M}_5$ exhibits therefore a delta like singularity at ${\cal M}_4$ and  satisfies
the following equations
$$\bar{\cal G}_{AB}={1\over6}\lambda^2\bar\gamma_{AB}+\kappa \bar{\cal T}_{AB}+\kappa\bar
T_{AB}\delta(y)\eqno(2.13)$$
where $\bar T_{AB}$ is the stress-energy tensor of matter on the brane. Integrating (13) (using (9))
across
$y=0$, yields the Lanczos-Darmois-Israel equations [1]
$$\bar T_{A5}=0\qquad,\qquad\kappa\bar T_{\mu\nu}=2{\cal L}_{\mu\nu}|_{y=0}=\lambda g_{\mu\nu}+\kappa
T_{\mu\nu} \eqno(2.14)$$
 which amount to identifying the tensor $T_{\mu\nu}$ we introduced with the stress-energy tensor of ordinary matter on
the brane. Equations (10) therefore become the equations for gravity in the brane.

As for the ``seed" tensor $\Theta_{\mu\nu}$, which is related to the $y$-derivative of the extrinsic curvature of ${\cal
M}_4$, it encapsulates the influence of the
geometry of the bulk near (rather than on) the brane, and can be expressed in terms of the Weyl tensor as in [11].

\bigskip

Now, it is clear that, if the geometry of the
bulk near the brane can be chosen at will, then the four dimensional Einstein equations
$$G_{\mu\nu}=8\pi G_N T_{\mu\nu}\eqno(2.15a)$$
with
$$8\pi G_N\equiv-{1\over6}\kappa\lambda\eqno(2.15b)$$ $G_N$ being Newton's
constant, can be exactly recovered on the brane. Indeed one simply has to impose
$${\cal T}_{5\mu}|_{y=0}=0\eqno(2.16a)$$
$$8{\cal T}_{55}|_{y=0}=\kappa\left({1\over3}T^2-T_{\rho\sigma}T^{\rho\sigma}\right)\eqno(2.16b)$$
$$2{\cal
T}_{\mu\nu}|_{y=0}-\Theta_{\mu\nu}=
\kappa\left[T_{\mu\rho}T^\rho_\nu-{1\over6}TT_{\mu\nu}+
{g_{\mu\nu}\over4}
\left(T_{\rho\sigma}T^{\rho\sigma}-{1\over3}T^2\right)\right]\,.\eqno(2.16c)$$
If one wishes however that ${\cal T}_{AB}$ describes some ``realistic" matter, then conditions (16) may not be fulfilled
for a realistic $T_{\mu\nu}$. Indeed, consider for example the case when matter in the bulk is a
massless scalar field $\Phi$ and the brane is a Robertson-Walker spacetime. Equation (16b) then reads
$$\psi^2+\dot\phi^2=-{\kappa\over6}\rho(\rho+3P)\eqno(2.17)$$
where $\psi\equiv{\partial\Phi\over\partial y}|_{y=0}$,  $\dot\phi\equiv{\partial\Phi\over\partial t}|_{y=0}$
 ($t$ being cosmic time), and
where $\rho$ and $P$ are the energy density and pressure of matter in the brane. For matter satisfying $P>-\rho/3$ (and
$\kappa\rho>0$) equation (17) has no solution.  (Defining $8\pi G_N\equiv-{\alpha\over6}\kappa\lambda$, $\alpha$
being an arbitrary constant, does not relax this constraint---nor the others we shall encounter.)

Let us summarize this section~: the metric near the brane is given in terms of the metric of the brane, the stress-energy
tensor of its  matter content and a ``seed" tensor by equation (7)~; gravity on the brane is described by equations
(10), ${\cal T}_{AB}$ being the stress-energy tensor of matter in the bulk~; these equations reduce to the four
dimensional Einstein equations if conditions (16) are satisfied.
\bigskip
\noindent
{\csc 3. The case of an Einstein bulk}
\bigskip

When ${\cal T}_{AB}|_{y=0}=0$
the bulk is an Einstein space near the brane. Introducing the ``seed" tensor
$$\Sigma_{\mu\nu}\equiv\Theta_{\mu\nu}+\kappa\left[T_{\mu\rho}T^\rho_\nu-
{1\over6}TT_{\mu\nu}+{g_{\mu\nu}\over4}
\left(T_{\rho\sigma}T^{\rho\sigma}-{1\over3}T^2\right)\right],\eqno(3.1)$$
the equations (2.10)  for gravity in the brane are then equivalent to
$$G_{\mu\nu}=8\pi G_N
T_{\mu\nu}-{\kappa\over2}\Sigma_{\mu\nu}\eqno(3.2a)$$
with the seed tensor $\Sigma_{\mu\nu}$  restricted to satisfy
$$\Sigma={\kappa\over2}\left({1\over3}T^2-T_{\rho\sigma}T^{\rho\sigma}
\right)\eqno(3.2b)$$
$$D_\mu\Sigma^\mu_\nu=0\,.\eqno(3.2c)$$
Note that equations (2b-c) define $\Sigma_{\mu\nu}$ (and hence $\Theta_{\mu\nu}$) up to a conserved and traceless
(that is radiation-like) tensor.
As for the metric near the brane, it is given by (2.7),
the tensor $\Theta_{\mu\nu}$ being constrained to satisfy conditions (2b-c) (with the definition (1)).

The Einstein equations will hold on the brane if, first
$$\Sigma_{\mu\nu}=0\quad\Longleftrightarrow\quad\Theta_{\mu\nu}=-\kappa\left[T_{\mu\rho}T^\rho_\nu-
{1\over6}TT_{\mu\nu}+{g_{\mu\nu}\over4}
\left(T_{\rho\sigma}T^{\rho\sigma}-{1\over3}T^2\right)\right],\eqno(3.3)$$
and if matter on the brane satisfies the constraint
$${1\over3}T^2-T_{\rho\sigma}T^{\rho\sigma}=0\,.\eqno(3.4)$$

Outside matter, $T_{\mu\nu}=0$. The constraint (4) is  hence satisfied, so that the Einstein
equations can hold on the brane if we choose $\Theta_{\mu\nu}=0$. The bulk metric near the brane  is then given by (2.7)
with $T_{\mu\nu}=\Theta_{\mu\nu}=0$, that is
$$\gamma_{\mu\nu}=g_{\mu\nu}\left(1+{1\over3}\lambda\,y+{1\over18}\lambda^2\,y^2+{\cal O}(y^3)\right)\eqno(3.5)$$
with $g_{\mu\nu}$ a Ricci flat metric. We recognize in (5) the expansion of $\gamma_{\mu\nu}=g_{\mu\nu}\exp
\lambda y/3$, the metric studied in [12], which is obtained by iteration of equations (2.8) when
${\cal T}_{AB}$ is imposed to be zero everywhere, and not only on the brane (see the Appendix). 

Inside matter, $T_{\mu\nu}\neq0$. However, at linear order in $T_{\mu\nu}$, the constraint (4) is still approximately
satisfied. Einstein's equations  can therefore still hold approximately on the brane if we choose
$\Theta_{\mu\nu}=0$. The bulk metric near the brane is given  in that case by (2.7) with
$\Theta_{\mu\nu}=0$ and the terms quadratic in $T_{\mu\nu}$ neglected.

Now, if terms quadratic in the stress-energy tensor cannot be neglected, then (4) becomes a
 restriction on the matter allowed on the brane. In the case of a perfect fluid~: $T_{\mu\nu}=(\rho+P)u_\mu u_\nu +
Pg_{\mu\nu}$ it yields
$$ P=-{1\over3}\rho\,.\eqno(3.6)$$

In conclusion, when the bulk is an Einstein space in the vicinity of the brane, the Einstein equations can be recovered on
the brane, at least at linear order in
$T_{\mu\nu}$,  by choosing
$\Theta_{\mu\nu}=0$, whatever the equation of state for the matter. However, when quadratic corrections are taken into
account, the equations for gravity on the brane differ from Einstein's, unless matter satisfies the condition (4) (or 6).

These results generalize known results which can be found in the literature when the brane is taken to be a
spatially flat Robertson-Walker spacetime [5-6].  Indeed in that case the metric $g_{\mu\nu}$
and the stress energy tensor $T_{\mu\nu}$ are supposed to be of the form
$$\eqalign{g_{tt}&=-1\quad,\quad\ \  g_{ti}=0\quad,\quad g_{ij}=a^2(t)\delta_{ij}\cr
T_{tt}&=\rho(t)\quad,\quad T_{ti}=0\quad,\quad T_{ij}=a^2P(t)\delta_{ij}\,.\cr}\eqno(3.7)$$
The solution of the equations for gravity on the brane is obtained by integrating either equations (3.2), or
equations (2.10) with
${\cal T}_{AB}=0$. Equation (2.10b) for example is the standard conservation law
$$\dot\rho+3{\dot a\over a}(\rho+P)=0\,.\eqno(3.8)$$
As for equation (2.10c) it reads
$${\ddot a\over a}+{\dot a^2\over
a^2}=-{\kappa\lambda\over36}(\rho-3P)-{\kappa^2\rho\over36}(\rho+3P)\eqno(3.9)$$
which  is equivalent, whatever the equation of state, to
$${\dot a^2\over a^2}={\kappa\over36}\rho(\kappa\rho-2\lambda)+{c\over a^4}\eqno(3.10)$$
with $c$ a constant of integration. We recognize in (10) the evolution equation for the scale factor $a$
first obtained in [5]. Finally equation (2.10a)  gives the ``seed" tensor $\Theta_{\mu\nu}$ (and hence the bulk metric
near the brane to second order in $y$, see equation (2.7)) as
$$\eqalign{\Theta_{00}&={\kappa\rho\over6}(5\rho+6P)-{6c\over\kappa a^4}\cr
\Theta_{ij}&=-a^2\delta_{ij}\left[{\kappa\over6}(3P^2+6P\rho+2\rho^2)+{2c\over\kappa a^4}\right]\,.\cr}\eqno(3.11)$$
In order, first, for these equations to reduce to the standard Friedmann equation, conditions (2.16) (with ${\cal
T}_{AB}=0$) must be fulfilled~: equation (2.16b)  implies that, as we have already seen, $P=-\rho/3$ (which renders
equation (9) linear in $\rho$)~; hence $\rho\propto a^{-2}$ ; 
equation (2.16c), together with (11) then imposes $c/a^4=-\kappa^2\rho^2/36$ (which
renders equation (10) equivalent to the Friedmann equation). 

Second, when terms quadratic in $T_{\mu\nu}$ can be
neglected, that is at late time, and when $\Theta_{\mu\nu}\approx0$, that is for $c=0$, then equation (10) tends, as
expected, to the Friedmann equation.

\bigskip
\noindent
{\csc 4. The case of an anti-de Sitter bulk}
\bigskip
Suppose now that the bulk is maximally symmetric near the brane, that is that its Riemann tensor ${\cal R}_{ABCD}$ is
such that
$${\cal R}_{ABCD}|_{y=0}=-{\lambda^2\over36}
(\gamma_{AC}\gamma_{BD}-\gamma_{AD}\gamma_{BC})|_{y=0}\,.\eqno(4.1)$$
Using again the standard four plus one decomposition (see equation (2.9)) as well as the quantities introduced in section
2 this equation can be rewritten as
$$\Theta_{\mu\nu}=-{\kappa\over2}\left[T_{\mu\rho}T^\rho_\nu+g_{\mu\nu}\left(T_{\rho\sigma}T^{\rho\sigma}
-{1\over3}T^2\right)\right]\eqno(4.2a)$$
$$0=D_\nu T_{\mu\rho}-D_\rho T_{\mu\nu}-{1\over3}(g_{\mu\rho}\partial_\nu T-g_{\mu\nu}\partial_\rho
T)\eqno(4.2b)$$
$$\eqalign{R_{\mu\nu\rho\sigma}&=-{\kappa T\over36}(2\lambda+\kappa
T)(g_{\mu\sigma}g_{\nu\rho}-g_{\mu\rho}g_{\nu\sigma})
-{\kappa^2\over4}(T_{\mu\sigma}T_{\nu\rho}-T_{\mu\rho}T_{\nu\sigma})\cr &+{\kappa\over12}(\lambda+\kappa
T)(g_{\mu\sigma}T_{\nu\rho}-g_{\mu\rho}T_{\nu\sigma}+
T_{\mu\sigma}g_{\nu\rho}-T_{\mu\rho}g_{\nu\sigma})\cr}\eqno(4.2c)$$ where $R_{\mu\nu\rho\sigma}$ is the Riemann
tensor of the brane metric $g_{\mu\nu}$. These equations which describe gravity on the brane  are more constraining
than equations (3.2). For example they imply that, outside  matter ($T_{\mu\nu}=0$)~: $R_{\mu\nu\rho\sigma}=0$
which means that the brane is necessarily flat (and not only a solution of (3.2) with $T_{\mu\nu}=0$ as is the case when
the bulk is only imposed to be an Einstein space). Outside matter we also have $\Theta_{\mu\nu}=0$ so that the metric
near the brane is, see equation (2.7)~:
$$\gamma_{\mu\nu}=\eta_{\mu\nu}\left(1+{\lambda\over3}y+{\lambda^2\over18}y^2+{\cal O}(y^3)\right)\eqno(4.3)$$
which is nothing but the lower order expansion of the anti-de Sitter metric in the Randall-Sundrum coordinates~:
$ds^2=dy^2+\eta_{\mu\nu}\exp(\lambda y/3)$, a metric which can be obtained by iteration of equations (2) when the
bulk is imposed to be anti-de Sitter spacetime everywhere an not only near the brane.

Inside matter, Equation (2a) gives $\Theta_{\mu\nu}$ in terms of the metric of the brane and its matter
content~; Equation (2b) is a constraint on the matter on the brane (which includes the conservation law $D_\mu
T^\mu_\nu=0$ but is in general more constraining than that)~; and Equation (2c) replaces the Einstein equations and
describes gravity on the brane.

As an example, consider a spatially flat Robertson-Walker brane where the metric $g_{\mu\nu}$
and the stress energy tensor $T_{\mu\nu}$ are given by (3.7).
For such an Ansatz equation (2b) turns out to be equivalent to the conservation law (3.8).
As for equation (2c) it is equivalent to (3.10) with
$$c=0\,.\eqno(4.4)$$
Finally (2a) gives $\Theta_{\mu\nu}$ as (3.11) with $c=0$, or, equivalently, the bulk metric near the brane to second
order in
$y$, which turns out to be the expansion at leading orders of the anti-de Sitter metric in the gaussian normal
coordinates introduced in [5]. Therefore, when one considers Robertson-Walker branes, the difference is quite tenuous
between imposing the bulk to be just an Einstein space or maximally symmetric near the brane~: in the latter case the
constant
$c$ must be zero~; in the former it is arbitrary and, when $P=-\rho/3$, can be chosen in such a way that the Friedmann
equations hold exactly. And in both cases, that is for $c=0$ or $c$ arbitrary, the terms quadratic in $T_{\mu\nu}$ become
negligible at  late time and the evolution of the brane tends to Friedmann's.

For less symmetric branes however, the Einstein equations cannot in general be recovered, even when terms quadratic in
$T_{\mu\nu}$ are negligible, as we shall now see.

In order to compare and contrast the brane gravity equations (2) with the four dimensional Einstein equations, let us
compute the brane Einstein tensor from equation (2c). We obtain
$$G_{\mu\nu}=-{\kappa\lambda\over6}T_{\mu\nu}+{\kappa^2\over4}
\left[-T_{\mu\rho}T^\rho_\nu+{1\over3}TT_{\mu\nu}+{1\over2}g_{\mu\nu}\left(T_{\rho\sigma}T^{\rho\sigma}
-{1\over3}T^2\right)\right]\eqno(4.5)$$
which, inside matter, can never exactly reduce to the Einstein equations (as we already saw in the case of a
Roberston-Walker brane). Now, at linear order in $T_{\mu\nu}$,  and with the identification $8\pi
G_N=-\kappa\lambda/6$, equations (5) do reduce to the four dimensional Einstein equations. However one must not
forget that they are not equivalent to the linear version of equations (2), that is
$$\Theta_{\mu\nu}\approx0\eqno(4.6a)$$
$$0\approx D_\nu T_{\mu\rho}-D_\rho T_{\mu\nu}-{1\over3}(g_{\mu\rho}\partial_\nu T-g_{\mu\nu}\partial_\rho
T)\eqno(4.6b)$$
$$ R_{\mu\nu\rho\sigma}\approx-{\kappa\lambda
\over18} T(g_{\mu\sigma}g_{\nu\rho}-g_{\mu\rho}g_{\nu\sigma}) +{\kappa\lambda\over12}
(g_{\mu\sigma}T_{\nu\rho}-g_{\mu\rho}T_{\nu\sigma}+
T_{\mu\sigma}g_{\nu\rho}-T_{\mu\rho}g_{\nu\sigma})\eqno(4.6c)$$ but only a consequence of those,  and one must
check that the chosen solution of Einstein's equations satisfies all equations (6). This is the case, as we have already
seen, when the brane is a Robertson-Walker spacetime. But consider now the case of an almost flat brane.

At zeroth order in $T_{\mu\nu}$ the brane must be flat and the metric can be taken
to be $g_{\mu\nu}=\eta_{\mu\nu}$. We decompose the stress-energy tensor at first order, as is usual, into
$$\eqalign{T_{00}=\rho\qquad,&\qquad T_{0i}= -\partial_iv- v_i\cr
T_{ij}= \delta_{ij}\left(P-{1\over3}\triangle
\Pi\right)+&\partial_{ij}\Pi+\partial_i\Pi_j+\partial_j\Pi_i+\Pi_{ij}\cr}\eqno(4.7)$$
where $\partial_i  v^i=\partial_i \Pi^i=\partial_i \Pi^{ij}= \Pi^i_i=0$, and where all components tend to zero at 
spatial infinity and at $t\to\pm\infty$. Equations (6b) then read
$$\eqalign{0&\approx \partial_i\left(\dot v+{2\over3}\rho+ P\right)+\dot v_i\cr
0&\approx
{1\over3}\delta_{ij}(\dot\rho-\triangle\dot\Pi)+\partial_{ij}(\dot\Pi+v)+\partial_i\dot\Pi_j+\partial_j\dot\Pi_i
+\partial_jv_i+\dot\Pi_{ij}\cr 0&\approx \partial_iv_j-\partial_jv_i\cr
0&\approx {1\over3}\delta_{jk}\partial_i(\rho-\triangle\Pi)-{1\over3}\delta_{ji}\partial_k(\rho-\triangle\Pi)+
\partial_j(\partial_i\Pi_k-\partial_k\Pi_i)+\partial_i\Pi_{jk}-\partial_k\Pi_{ij}\cr}\eqno(4.8)$$
which include the conservation laws $\partial_\mu T^\mu_\nu\approx 0$, that is
$$\dot\rho+\triangle v\approx 0\qquad,\qquad \dot
v+P+{2\over3}\triangle\Pi\approx 0\qquad,\qquad\dot v_i+\triangle\Pi_i\approx 0\eqno(4.9)$$
but also impose matter to obey the following, very contrived, equation of state
$$\rho\approx \triangle\Pi\quad,\quad v\approx -\dot\Pi\quad,\quad
P\approx \ddot\Pi-{2\over3}\triangle\Pi\qquad,\qquad v_i\approx \Pi_i\approx \Pi_{ij}\approx 0\,.\eqno(4.10)$$

Matter being described solely in terms of the anisotropic stress $\Pi$ by equations (9-10), equation (6c) gives the
Riemann tensor of the brane as
$$R_{\mu\nu\rho\sigma}\approx {\kappa\lambda\over12}(\eta_{\mu\sigma}\partial^2_{\nu\rho}+
\eta_{\nu\rho}\partial^2_{\mu\sigma}-\eta_{\mu\rho}\partial^2_{\nu\sigma}
-\eta_{\nu\sigma}\partial^2_{\mu\rho})\Pi\eqno(4.11)$$
which defines uniquely the geometry of the brane as the conformally flat metric
$$g_{\mu\nu}\approx \left(1+{1\over6}\kappa\lambda\,\Pi\right)\eta_{\mu\nu}\,.\eqno(4.12)$$
Finally equation (2.7), together with (6a) and (9-10) gives the metric near the brane as
$$\gamma_{\mu\nu}\approx  \eta_{\mu\nu}\left(1+{1\over3}\lambda\,y+{1\over18}\lambda^2\,y^2\right)-
\kappa y\left(1+{1\over6}\lambda\,y\right)\partial_{\mu\nu}\Pi+{\cal O}(y^3)\,.\eqno(4.13)$$
This metric describes, by construction, a strictly anti-de Sitter bulk near the brane and can therefore be cast into the
form (3) by a mere change of coordinates. However this change of coordinates changes the equation giving the position
of the brane which is no longer given by $y=0$. The last term hence describes, in gaussian normal coordinates, the
so-called ``brane-bending" effect (see also [13]).

\bigskip
\noindent
{\csc 5. Conclusions}
\bigskip
The main result of this paper is that the question of whether or not Einstein's equations are recovered on a brane depends
crucially on the geometry of the bulk near the brane. If the bulk is an Einstein space near the brane, then the Einstein
equations, at least at linear order in the stress-energy tensor
$T_{\mu\nu}$,  can be recovered, {\it whatever} the equation of state of the matter on the brane. However, when
quadratic terms in $T_{\mu\nu}$ cannot be neglected,  Einstein's equations hold
only if the equation of state for matter is
$P=-\rho/3$ (for a perfect fluid). If, now, the bulk is imposed to be strictly anti-de Sitter space near the brane, then
the brane {\it must} be flat outside matter. Moreover the Einstein equations can never be recovered when terms quadratic
in
$T_{\mu\nu}$ are important. Finally, when terms quadratic in
$T_{\mu\nu}$ can be neglected, the linearized Einstein equations can hold on a quasi-minkowskian brane, but only for
very contrived matter.

This last result does not by any means imply that the linearized Einstein equations cannot be recovered in the
Randall-Sundrum scenario. Indeed, in that scenario, the bulk is a {\it perturbed} anti-de Sitter space, that is an Einstein
space.  The results of section 3 then tell us that if we choose, at zeroth order in $\lambda T_{\mu\nu}$ and
$\Theta_{\mu\nu}$ the flat solution of the brane equation for gravity (3.2a), then, at linear order, gravity on the brane
is governed by the equation
$$\tilde G_{\mu\nu}\approx8\pi G_N T_{\mu\nu}-{\kappa\over2} \Theta_{\mu\nu}\eqno(5.1)$$
where $\tilde G_{\mu\nu}$ is the Einstein tensor of the metric $g_{\mu\nu}=\eta_{\mu\nu}+h_{\mu\nu}$ at linear order
in $h_{\mu\nu}$ and where $\kappa \Theta_{\mu\nu}$ describes a radiation-like fluid
which, a priori, can contribute as much as $G_N T_{\mu\nu}$ to $\tilde G_{\mu\nu}$ but which can also be chosen to be
zero. Finally the metric near the brane reads
$$\eqalign{\gamma_{\mu\nu}\approx  \eta_{\mu\nu}&\left(1+{1\over3}\lambda\,y+{1\over18}\lambda^2\,y^2\right)-
\kappa y\left(1+{1\over6}\lambda\,y\right)\left(T_{\mu\nu}-{1\over3}T\eta_{\mu\nu}\right)\cr&-
{1\over2}\kappa y^2\left(\Theta_{\mu\nu}-{1\over3}\Theta\eta_{\mu\nu}\right)+{\cal O}(y^3)\cr}\eqno(5.2)$$
which is not simply, as (4.13), anti-de Sitter metric in disguise.

We leave to another work [10] the comparison of the ``brane world" point of view developped here with the ``bulk" point of
view where the bulk is imposed to be a perturbed anti-de Sitter space {\it everywhere} and not only near the brane and
where the perturbations are imposed to satisfy boundary conditions which may restrict $\Theta_{\mu\nu}$
and/or $T_{\mu\nu}$.
\bigskip
\bigskip
\noindent
{\csc Acknowledgements} 
\bigskip
We thank David Langlois and Tom\'a\v s Dole\v zel for fruitful discussions.
\bigskip
\vfill\eject
\bigskip
\noindent
{\csc Appendix}
\bigskip
Consider a five dimensional spacetime ${\cal V}_5$ in gaussian normal coordinates $x^A=(x^\rho,y)$
$$ds^2=\gamma_{AB}dx^A dx^B=\epsilon dy^2+\gamma_{\mu\nu}(x^\rho,y)dx^\mu dx^\nu\eqno(B1)$$
with $\epsilon=\pm1$ and expand the metric coefficients $\gamma_{\mu\nu}(x^\rho,y)$ near the surface $y=0$ as
$$\gamma_{\mu\nu}(x^\rho,y)=g_{\mu\nu}(x^\rho)+k_{\mu\nu}(x^\rho)\,y+{1\over2}l_{\mu\nu}(x^\rho)\,y^2
+{1\over6}m_{\mu\nu}(x^\rho)\,y^3+{\cal O}(y^4)\,.\eqno(B2)$$
The extrinsic curvature of the surface $y=Const.$ and its $y$-derivative are given by
$$\eqalign{{\cal K}_{\mu\nu}&\equiv-{1\over2}{\partial\gamma_{\mu\nu}\over\partial y}=
-{1\over2}\left(k_{\mu\nu}+l_{\mu\nu}\,y+{1\over2}m_{\mu\nu}\,y^2\right)+{\cal O}(y^3)\cr
{\partial{\cal K}_{\mu\nu}\over\partial y}&=-{1\over2}\left(l_{\mu\nu}+m_{\mu\nu}\,y\right)+{\cal
O}(y^2)\,.\cr}\eqno(B3)$$
The Riemann tensor of the metric (B1) reads
$$\eqalign{{\cal R}_{y\mu y\nu}&={\partial\over\partial y}{\cal K}_{\mu\nu}+{\cal K}_{\rho\mu}{\cal K}^\rho_\nu\cr
{\cal R}_{y\mu \nu\rho}&=\nabla_\nu{\cal K}_{\mu\rho}-\nabla_\rho{\cal K}_{\mu\nu}\cr
{\cal R}_{\mu \nu\rho\sigma}&=^4{\cal R}_{\mu \nu\rho\sigma} +\epsilon\left({\cal K}_{\mu\sigma}{\cal K}_{\nu\rho}
-{\cal K}_{\mu\rho}{\cal K}_{\nu\sigma}\right)\cr}\eqno(B4)$$
where $\nabla_\mu$ and $^4{\cal R}_{\mu \nu\rho\sigma}$ are the covariant derivative and Riemann tensor associated
with the metric $\gamma_{\mu\nu}(x^\rho,y)|_{y=Const.}$. Expanding (B4) to first order in $y$, it is a straightforward
calculation to obtain the Einstein tensor of the metric (B1) as 
$$\eqalign{{\cal G}_{yy}=&-{\epsilon\over2}R+{1\over8}(k^2-k\,. k)\cr &+{1\over4}\,y\left[2\epsilon(\square
k- D\,. k+ k\,. R)+kl-k\,. l- k(k\,. k)+k\,. k\,. k\right]+{\cal O}(y^2)\cr}\eqno(B5a)$$
where $R$ and $D$ are the scalar curvature and covariant derivative of the metric $g_{\mu\nu}$, where traces are
defined by means of
$g^{\mu\nu}$, where $\square\equiv D_\rho D^\rho$ and where $a\,.b\equiv a_{\mu\nu}b^{\mu\nu}$,
$a\,.b\,.c\equiv a_{\mu\nu}b^{\nu\rho}c_\rho^\mu\,$ ;
$$\eqalign{{\cal G}_{y\mu}&={1\over2}\left(D_\nu k^\nu_\mu-\partial_\mu k\right)\cr&
-{1\over2}y\left[\partial_\mu l-D_\nu
l^\nu_\mu-{1\over2}k^\nu_\mu\partial_\nu k+D_\nu(k^{\nu\rho}k_{\rho\mu})
-{3\over4}\partial_\mu(k\,. k)\right]+{\cal O}(y^2)\,,
\cr}\eqno(B5b)$$
Finally
$$\eqalign{{\cal
G}_{\mu\nu}&=G_{\mu\nu}+{\epsilon\over2}\left[g_{\mu\nu}l-l_{\mu\nu}+k^\rho_\mu k_{\rho\nu}
-{1\over2}kk_{\mu\nu}+{1\over4}g_{\mu\nu}(k^2-3k\,.k)\right]
\cr&
+{\epsilon\over2}y\left[mg_{\mu\nu}-m_{\mu\nu}+k_{\mu\rho}l^\rho_\nu+k_{\nu\rho}l^\rho_\mu-kl_{\mu\nu}
-{1\over2}lk_{\mu\nu}+{1\over2}g_{\mu\nu}(kl-2k\,.l)\right]\cr&
-{\epsilon\over2}y\left\{ k_{\mu\rho}k^{\rho\lambda}k_{\lambda\nu}+{1\over4}(k\,.k-k^2)k_{\mu\nu}
+{1\over2}[k(k\,.k)-3k\,.k\,.k]\right\}\cr&
+{1\over2}y\left[D_{\rho\mu}k^\rho_\nu+D_{\rho\nu}k^\rho_\mu-\square k_{\mu\nu}-D_{\mu\nu}k
-g_{\mu\nu}\left(D\,.k-\square k-k\,. R\right)-Rk_{\mu\nu}\right]\cr}\eqno(B5c)$$
where $G_{\mu\nu}$ is the Einstein tensor of the metric $g_{\mu\nu}$.

If now ${\cal V}_5$ is imposed to be an Einstein space, ${\cal G}_{AB}=\Lambda\gamma_{AB}$, not only on the
brane as in the main text but at linear order in $y$, then
$${\cal G}_{yy}=\epsilon\Lambda\quad,\quad
{\cal G}_{y\mu}=0\quad,\quad
{\cal G}_{\mu\nu}=\Lambda(g_{\mu\nu}+y\,k_{\mu\nu})+{\cal O}(y^2)\,.\eqno(B6)$$

Suppose now that we are given a metric and its $y$-derivative at $y=0$, that is that we know $g_{\mu\nu}$ and
$k_{\mu\nu}$ satisfying the constraints
$$D_\nu k^\nu_\mu-\partial_\mu k=0\quad,\quad -\epsilon R+{1\over4}(k^2-k\,.k)=2\epsilon\Lambda\,.\eqno(B7)$$
Then equations (B5a-b) with (B6) are satisfied at zeroth order in $y$, and equation (B5c) (with (B6)) gives $l_{\mu\nu}$
in terms of $g_{\mu\nu}$ and $k_{\mu\nu}$ as
$$l_{\mu\nu}-g_{\mu\nu}l=2\epsilon G_{\mu\nu}+k^\rho_\mu k_{\rho\nu}-{1\over2}kk_{\mu\nu}
+{1\over4}g_{\mu\nu}(k^2-3k\,. k)-2\epsilon\Lambda g_{\mu\nu}\,.\eqno(B8)$$ Hence the zeroth order Einstein
equations give us the metric near the brane at quadratic order in $y$.
 
It is then straightforward to see that equations (B5a-b) together with (B6) are satisfied at linear order in $y$. As for
equation (B5c) together with (B6) it gives $m_{\mu\nu}$, and hence the metric at cubic order in $y$. 

Iterating this procedure, assuming that  ${\cal V}_5$ is an Einstein space up to higher and higher order in $y$ should
give the metric of ${\cal V}_5$  everywhere (or at least in a finite region near $y=0$).

To make the connection with the main text, first take $\epsilon=+1$ and $\Lambda=\lambda^2/6$ and introduce the
``$\tau$"-tensor
$${\kappa\over2}\tau_{\mu\nu}\equiv{\cal K}_{\mu\nu}-\gamma_{\mu\nu}{\cal
K}-{1\over2}\lambda\gamma_{\mu\nu}\eqno(B9)$$
and expand it as
$$\tau_{\mu\nu}=T_{\mu\nu}+y\,\Theta_{\mu\nu}+{1\over2}y^2{\cal H}_{\mu\nu}+{\cal O}(y^3)\,.\eqno(B10)$$
Using (B2) we have
$$\eqalign{{\kappa\over2}T_{\mu\nu}&=-{1\over2}[k_{\mu\nu}+(\lambda-k)g_{\mu\nu}]\cr
{\kappa\over2}\Theta_{\mu\nu}&=-{1\over2}[l_{\mu\nu}+(\lambda-k)k_{\mu\nu}-(l-k\,.k)g_{\mu\nu}]\cr
{\kappa\over2}{\cal H}_{\mu\nu}&=-{1\over2}[m_{\mu\nu}+(\lambda-k)l_{\mu\nu}-2(l-k\,.k)k_{\mu\nu}
-(m-3k\,.l+2k\,.k\,.k)g_{\mu\nu}]\,.\cr}\eqno(B11)$$

Consider now  the particularly simple example where, instead of knowing $g_{\mu\nu}$ and $k_{\mu\nu}$ (or
equivalently $T_{\mu\nu}$), we are given
$$T_{\mu\nu}=\Theta_{\mu\nu}=0\,.\eqno(B12)$$
Then we first get from (B11)
$$k_{\mu\nu}={\lambda\over3}g_{\mu\nu}\quad\hbox{and}\quad
l_{\mu\nu}={\lambda^2\over9}g_{\mu\nu}\,.\eqno(B13)$$
Equations (B5) together with (B6) then give, at zeroth order in $y$
$$G_{\mu\nu}=0\eqno(B14)$$
and, at linear order in $y$
$$m_{\mu\nu}={\lambda^3\over27}g_{\mu\nu}\quad\hbox{and}\quad {\cal H}_{\mu\nu}=0\,.\eqno(B15)$$
The metric near the brane is then the expansion, up to cubic order in $y$ of the metric
$\gamma_{\mu\nu}=g_{\mu\nu}\exp (\lambda y/3)$, $g_{\mu\nu}$ being a Ricci flat metric. We also have that
$\tau_{\mu\nu}$ is zero up to quadratic order in $y$. Iterating the procedure, with the condition that ${\cal V}_5$ is an
Einstein space everywhere would yield (at least in a finite region near $y=0$)
$$ G_{\mu\nu}=0\quad,\quad\gamma_{\mu\nu}=g_{\mu\nu}\exp (\lambda y/3)\quad,\quad \tau_{\mu\nu}=0\eqno(B16)$$
that is the metric discussed in [12].
\bigskip\bigskip
\noindent
{\csc References}
\bigskip
\item{[1]} W. Israel, Nuovo Cimento B44, 1, (1966), Nuovo Cimento B48, 463 (1967); K. Lanczos, Phys. Z. 23, 539 (1922)
and Ann. Phys. (Leipzig), 74, 518 (1924); G. Darmois, M\'emorial des Sciences Math\'ematiques XXV, Gauthier-Villars,
Paris (1927)

\item{[2]} I. Antoniadis, N. Arkani-Hamed, S. Dimopoulos, G. Dvali, Phys. Lett. B436, 257 (1998);  P.
Horava, E. Witten, Nucl. Phys. B460, 506 (1996), Nucl. Phys. B475, 94 (1996)

\item{[3]} L. Randall, R. Sundrum, Phys. Rev. Letters 83, 4690 (1999)

\item{[4]} J. Garriga, T. Tanaka,  Phys. Rev. Letters 84 (2000) 2778;
 C. Csaki, J. Erlich, T.J. Hollowood, Y. Shirman, Nucl. Phys. B581 (2000) 309; S.B. Giddings, E. Katz, L. Randall,
 JHEP 0003 (2000) 023;
C. Csaki, J. Erlich, T.J. Hollowood, Phys. Rev. Lett. 84 (2000) 5932;
 C. Csaki, J. Erlich, T.J. Hollowood, Phys. Lett. B481 (2000) 107;
  I.Ya. Aref'eva, M.G. Ivanov, W. Muck, K.S. Viswanathan, I.V.
Volovich, Nucl. Phys. B590 (2000) 273;  Z. Kakushadze, Phys. Lett. B497 (2000) 125

\item{[5]} P. Bin\'etruy, C. Deffayet, U. Ellwanger,  D. Langlois,  Phys Lett. B477 (2000) 285 

\item{[6]} P. Kraus, JHEP 9912 (1999) 011; S. Mukohyama,  Phys. Lett. B473 (2000) 241;
 D. N. Vollick, Class. Quant. Grav. 18 (2001) 1; D. Ida, JHEP 0009 (2000) 014; 
S. Mukohyama, T. Shiromizu, K. Maeda, Phys. Rev. D62 (2000) 024028; 
 R. Maartens, Phys. Rev. D62 (2000) 084023; P. Kanti, I. I. Kogan, K. A. Olive, M. Pospelov, Phys. Lett. B468 (1999) 31

\item{[7]} S. Mukohyama, Phys. Rev. D62 (2000) 084015;
 H. Kodama, A. Ishibashi and O. Seto, Phys. Rev. D62 (2000) 064022;
 D. Langlois, Phys. Rev. D62 (2000) 126012;
  C. van de
Bruck, M. Dorca, R. Brandenberger, A. Lukas, Phys. Rev. D62 (2000) 123515;
 K. Koyama, J. Soda, Phys. Rev. D62 (2000) 123502;
 D. Langlois, R. Maartens, D. Wands. Phys. Lett. B489 (2000) 259;
 S. Mukohyama, Class. and Quant. Grav. 17 (2000) 4777;
C. Gordon, D. Wands, B. Bassett, R. Maartens, Phys. Rev. D63 (2001) 023506;
C. Gordon. R. Maartens, Phys. Rev. D63 (2001) 044022;
D. Langlois, R. Maartens, M. Sasaki, D. Wands, Phys. Rev. D63 (2001) 023506;
N. Deruelle, T. Dolezel, J. Katz, ``Perturbations of brane world", hep-th/0010215;
C. van de Bruck, M. Dorca, ``On cosmological perturbations on a brane in an anti-de Sitter bulk", hep-th/0012073

\item{[8]} P. Bin\'etruy, C. Deffayet, D. Langlois, ``The radion in brane cosmology", hep-th/\-0101243 and references
therein

\item{[9]} I. Low, A. Zee, Nucl. Phys. B585 (2000) 1395;
 N. Deruelle and T. Dolezel, Phys. Rev. D62 (2000) 103502;
I.P. Neupane, JHEP 0009 (2000) 1040;
J.E. Kim and H. M. Lee, ``Gravity in the Einstein-Gauss-Bonnet theory with the Randall-Sundrum background", hep-th/0011118;
O. Corradini and Z. Kakushadze, Phys. Lett. B494 (2000) 302

\item{[10]} N. Deruelle, T. Dolezel, J. Katz ``On linearized gravity in the Randall-Sundrum scenario", in preparation

\item{[11]} T. Shiromizu, K. Maeda, M. Sasaki, Phys. Rev. D62 (2000) 024012; 
 M. Sasaki, T, Shiromizu, K. Maeda, Phys. Rev D62 (2000) 024008;
R. Maartens, ``Geometry and dynamics of brane worlds", gr-qc/0101059

\item{[12]} A. Chamblin, S.W. Hawking, H.S. Reall, Phys. Rev. D61 (2000) 065007

\item{[13]} M. Dorca, C. van de Bruck, ``Cosmological perturbations in brane worlds : brane bending and anisotropic
stresses", hep-th/0012116

\end